\RequirePackage{tikz}
\usetikzlibrary{
	positioning,      calc, arrows,
	arrows.meta,      fit,  shapes,
	shapes.multipart, snakes
}

\documentclass[sn-basic]{sn-jnl}

\normalbaroutside

\usepackage[utf8]{inputenc}
\usepackage[english]{babel}
\usepackage{graphicx,float,subcaption,placeins}
\usepackage[algo2e]{algorithm2e} 
\usepackage{amsmath,amssymb,amsthm,bbm}
\usepackage{bm}
\usepackage{braket}
\usepackage{csquotes}
\usepackage{mathtools}
\usepackage{natbib}
\usepackage{subcaption}
\usepackage{siunitx}
\usepackage{xspace}
\usepackage{nicefrac}
\usepackage[normalem]{ulem}
\usepackage[mode=buildnew]{standalone}

\usepackage{soulpos}
\usepackage[capitalize]{cleveref}

\usepackage{tikz}
\usepackage{pgfplots,pgfplotstable}
\pgfplotsset{compat=newest}
\usetikzlibrary{calc,math,decorations.pathreplacing,decorations.pathmorphing}
\usepgfplotslibrary{groupplots,fillbetween}
\usepackage{siunitx}

\newcommand{\eg}{e.\,g.\@\xspace}

\newcommand{\ie}{i.\,e.\@\xspace}
\newcommand{\bC}{\boldsymbol{C}}
\newcommand{\bI}{\boldsymbol{I}}
\newcommand{\bmu}{\boldsymbol{\mu}}

\newcommand{\bQ}{\boldsymbol{Q}}
\newcommand{\bR}{\boldsymbol{R}}
\newcommand{\bsigma}{\boldsymbol{\sigma}}
\newcommand{\bSigma}{\boldsymbol{\Sigma}}
\newcommand{\btheta}{\boldsymbol{\theta}}
\newcommand{\bw}{\boldsymbol{w}}
\newcommand{\bx}{\boldsymbol{x}}
\newcommand{\bb}{\boldsymbol{b}}
\newcommand{\cB}{\mathcal{B}}
\newcommand{\cD}{\mathcal{D}}
\newcommand{\cI}{\mathcal{I}}

\newcommand{\cO}{\mathcal{O}}

\newcommand{\indicator}[1]{\mathbbm{1}_{\lbrace#1 \rbrace}}

\newcommand{\subjectto}{\text{s.\,t.}}
\DeclarePairedDelimiter\norm{\lVert}{\rVert}%

\colorlet{smcolor}{blue!30}
\colorlet{rhcolor}{red!30}
\colorlet{sacolor}{green!30}
\colorlet{mwcolor}{orange!30}
\colorlet{npcolor}{yellow!30}
\colorlet{commentcolor}{gray!50}

\jyear{2022}%

\theoremstyle{thmstyleone}%
\newtheorem{theorem}{Theorem}
\newtheorem{proposition}[theorem]{Proposition}%

\theoremstyle{thmstyletwo}%

\theoremstyle{thmstylethree}%

\begin{document}

\title[Feature Selection on Quantum Computers]{Feature Selection on Quantum Computers}

\author*[1]{\fnm{Sascha} \sur{M\"{u}cke}}\email{sascha.muecke@tu-dortmund.de}

\author[2]{\fnm{Raoul} \sur{Heese}}

\author[2]{\fnm{Sabine} \sur{M\"{u}ller}}

\author[3]{\fnm{Moritz} \sur{Wolter}}

\author[4]{\fnm{Nico} \sur{Piatkowski}}

\affil*[1]{\orgname{TU Dortmund}, \orgdiv{AI Group}, \orgaddress{\city{Dortmund}, \postcode{44227}, \country{Germany}}}

\affil[2]{\orgname{Fraunhofer FZML, ITWM}, \orgaddress{\city{Kaiserslautern}, \postcode{67663}, \country{Germany}}}

\affil[3]{\orgname{Fraunhofer FZML, SCAI}, \orgaddress{\city{Sankt Augustin}, \postcode{53757}, \country{Germany}}}

\affil[4]{\orgname{Fraunhofer IAIS}, \orgaddress{\city{Sankt Augustin}, \postcode{53757}, \country{Germany}}}

\abstract{
In machine learning, fewer features reduce model complexity.
Carefully assessing the influence of each input feature on the model quality is therefore a crucial preprocessing step.
We propose a novel feature selection algorithm based on a quadratic unconstrained binary optimization (QUBO) problem, which allows to select a specified number of features based on their importance and redundancy.
In contrast to iterative or greedy methods, our direct approach yields higher-quality solutions.  
QUBO problems are particularly interesting because they can be solved on quantum hardware.
To evaluate our proposed algorithm, we conduct a series of numerical experiments using a classical computer, a quantum gate computer and a quantum annealer.
Our evaluation compares our method to a range of standard methods on various benchmark datasets. We observe competitive performance.

}


\maketitle

\section{Introduction}
\label{sec:intro}

Machine learning (ML) models excel in a wide range of data analytics tasks, such as classification, regression, and data generation.
However, most model families scale with the number of input data dimensions, \ie, with the number of features.
That is, a model with more features requires more memory and more computational effort for its training. 
Small and fast models are key for applications with tight resource constraints, \eg, in embedded systems.
Therefore, it is a common goal in ML pipelines to reduce the number of features with minimum information loss by a suitable transformation from raw data to training data, a strategy also known as dimensionality reduction \citep{van_der_maaten_dimensionality_2009}.\par
An important example for such a strategy is feature selection (FS), where the input dimension is reduced by selecting only a subset of all available features without performing additional transformations \citep{chandrashekar_survey_2014}.
This approach is especially effective when dealing with data from sources that produce many redundant or irrelevant features, which can be eliminated without significantly impacting the output quality.
Consider, for example, trying to diagnose a specific disease from a vast array of medical measurements such as body temperature, concentration of various substances in a patient's blood or heart rate.
Reducing the number of features that are necessary for this diagnosis not only allows for smaller models but might even help experts pin down what causes the disease.
Consequently, FS can be seen as a tool to both reduce model complexity as well as to improve ML interpretability.

The main contribution of this paper consists of two parts. First, we propose a novel FS algorithm for selecting a specific number of features using a quadratic \emph{unconstrained} binary optimization formulation that can be applied to any combination of redundancy and importance measures. And second, we benchmark our proposed algorithm on different data sets using both classical and actual quantum hardware to demonstrate its effectiveness.

The remaining paper is organized as follows. First, in \cref{sec:method}, we introduce our FS algorithm. Subsequently, we discuss in \cref{sec:qubo} how unconstrained binary optimization problems, on which our FS algorithm is based, can be solved. We relate our contribution to previous research results in \cref{sec:related}. In \cref{sec:experiments}, we perform and evaluate a series of numerical experiments. Finally, we close with a conclusion in \cref{sec:conclusion}.

\section{Method}
\label{sec:method}

In the present section, we describe our proposed FS algorithm using a quadratic unconstrained binary optimization (QUBO) problem, which can be solved either by classical methods or with quantum computing. We start with a general problem definition and subsequently prove that the number of features can be selected by tuning a user-defined weighting parameter. Based on these prerequisites, we present our FS algorithm.

\subsection{QUBO feature selection}
\label{sec:qfs}
Presume a classification task on a data set $\cD:=\lbrace(\bx^i,y^i)\rbrace_{i\in [N]}$ with $n$-dimensional features $\bx^i \in \mathcal{X} \subseteq \mathbb{R}^n$ and class labels $y^i \in \mathcal{Y} \subseteq \mathbb{N}$ for all $i\in[N]$, where $[N]$ denotes the set $\lbrace 1,\dots,N\rbrace$.
The problem of FS corresponds to finding a subset $S\subset [n]$ of these $n$ features, such that the reduced data set $\cD_S=\lbrace (\bx^i_S,y^i)\rbrace_{i\in [n]}$ with $\bx_S:=(x_j)_{j\in S}$ leads to comparable performance as the original data for some data-driven task, such as classification.
Typically, this subset is found by solving a suitably posed optimization problem, which can also explicitly depend on the classification model.\par

We propose a model-independent formulation
\begin{equation}\label{eq:qfs}
	\bx^* := \underset{\bx\in\lbrace 0,1\rbrace^n}{\arg\min} Q(\bx,\alpha)
\end{equation}
to obtain the selected features $\bx^* \in \mathcal{X}^* \subset \mathcal{X}$. We represent the subset $S$ as a binary indicator vector $\bx:=(x_1,\dots,x_n)\in\lbrace 0,1\rbrace^n$, such that $i\in S$ if and only if $x_i=1$ for all $i \in [n]$.
The objective function reads
\begin{equation}\label{eq:Q}
	Q(\bx,\alpha) := - \alpha \sum_{i=1}^n I_i x_i + (1-\alpha) \sum_{i,j=1}^n R_{ij} x_i x_j,
\end{equation}
where the user-defined parameter $\alpha \in [0,1]$ balances the influence of the two terms, which we specify in the following as importance term and redundancy term.
The importance term contains the elements
\begin{equation}\label{eq:Ii}
	I_i := I(x_i;y) \geq 0
\end{equation}
of the importance vector $\bI\in\mathbb{R}_{0+}^{n}$. The importance vector represents the mutual information $I(x_i;y)$ of the individual features $x_1,\dots,x_n$ with the class label $y$ and is therefore a measure for the importance of each feature.
In our objective function, the importance is maximized.
Furthermore, the redundancy term contains the elements
\begin{equation}\label{eq:Rij}
	R_{ij} := I(x_i;x_j) \geq 0
\end{equation}
of the pairwise redundancy matrix $\bR\in\mathbb{R}^{n \times n}$, which by definition is symmetric and positive semidefinite.
This matrix represents the mutual information (MI) $I(x_i;x_j)$ among the individual features and therefore measures their redundancy.
For $i=j$, we set $R_{ii}=0$, since a feature is not redundant by itself.
In our objective function, the redundancy is minimized.

The calculation of mutual information requires explicit knowledge about the joint probability mass function of features and labels and the corresponding marginals, which are in general difficult to estimate empirically for real-valued data.
Therefore, we map all available feature values from the data set $\cD$ into $B$ discrete bins.
Specifically, for each separate feature dimension $i$ we take all $\nicefrac{\ell}{B+1}$-quantiles for $\ell\in\lbrace 0,\dots,B\rbrace$, which we denote by $q^\ell_i$.
With these, we define bins $\cB^\ell_i$ as intervals $[q^{\ell-1}_i,q^{\ell}_i)$ for $\ell\in[B-1]$, and $\cB^B_i=[q^{B-1}_i,q^B_i]$.
Finally, we set $b_i^j:=\ell$ for the single $\ell$ that fulfills $x_i^j\in\cB^\ell_i$.
Since the labels are discrete by definition, no binning is necessary in $\mathcal{Y}$.
This way, we obtain a discretized data set $\hat{\cD}=\lbrace (\bb^i,y^i)\rbrace_{i\in[N]}$ with $b^j_i\in [B]$ for all $j\in[N]$ and $i\in[n]$.

The empirical probability mass function after discretization reads
\begin{align}
	\hat{p}(\bb,y) := \frac{1}{N}\sum_{(\bb',y')\in\hat{\cD}}\indicator{\bb=\bb'\wedge y=y'}
\end{align}
with the indicator function
\begin{align}
	\indicator{P} := \begin{cases} 1 & \text{if $P$ evaluates to true} \\ 0 & \text{otherwise} \end{cases}
\end{align}
defined for logical statements $P$.
Consequently, we can approximate the information entropies in \cref{eq:Ii,eq:Rij} as
\begin{equation}\label{eq:lmi}
	I(x_i,y) \approx \sum_{b\in[B]}\sum_{y\in\mathcal{Y}}\hat{p}_{X_i,Y}(b,y)\log\left(\frac{\hat{p}_{X_i,Y}(b,y)}{\hat{p}_{X_i}(b)\hat{p}_{Y}(y)}\right)
\end{equation}
and
\begin{equation}\label{eq:Ii:approx}
	I(x_i,x_j) \approx \sum_{b\in[B]} \sum_{b'\in[B]}\hat{p}_{X_i,X_j}(b,b')\log\left(\frac{\hat{p}_{X_i,X_j}(b,b')}{\hat{p}_{X_i}(b)\hat{p}_{X_j}(b')}\right),
\end{equation}
respectively, where we make use of the marginals
\begin{align}
	\hat{p}_{X_i,X_j}(b_i,b_j) := \sum_{y\in\mathcal{Y}, b_k\in[B] \forall k \neq i,j} \hat{p}(\bb,y),
\end{align}
\begin{align}
	\hat{p}_{X_i,Y}(b_i,y) := \sum_{b_k\in[B]\forall k \neq i} \hat{p}(\bb,y),
\end{align}
\begin{align}
	\hat{p}_{X_i}(b_i) := \sum_{y\in\mathcal{Y}, b_k\in[B] \forall k \neq i} \hat{p}(\bb,y),
\end{align}
and
\begin{align}
	\hat{p}_{Y}(y) := \sum_{b_k\in[B] \forall k} \hat{p}(\bb,y)
\end{align}
for the probability mass functions of features subsets and labels.
Discretization allows us to approximate the MI values while greatly simplifying the estimation procedure, since no assumption on the underlying probability distribution of the data is required.
Moreover, the estimation is consistent, \ie, when the number of bins approaches infinity, we will recover the true MI between continuous features \cite[Theorem 3.2]{mandros_discovering_2020}.
To simplify our notation, we omit the dependence of $R_{ij}$ and $I_{i}$ on $B$.

\begin{figure*}
	\centering
	\includegraphics[width=\textwidth]{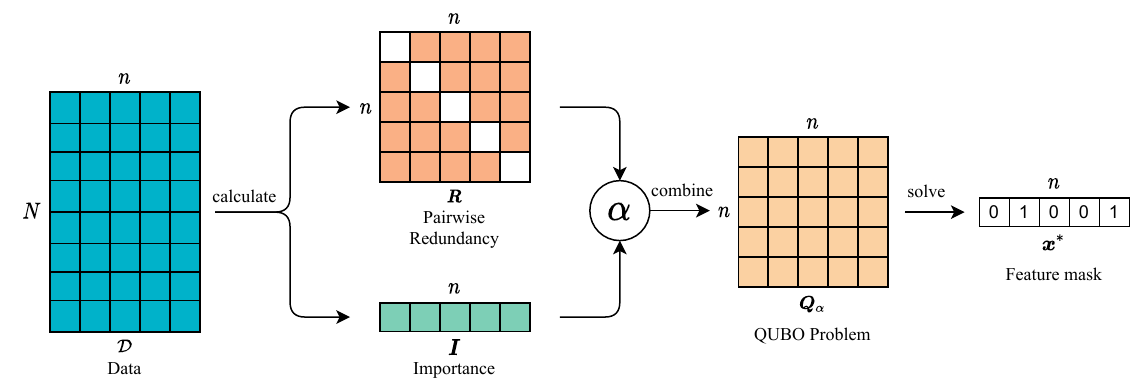}
	\caption{Our proposed QFS pipeline: From a given data set, the redundancy matrix $\bR$ and the importance vector $\bI$ are calculated, \cref{eq:Ii} and \cref{eq:Rij}.
		They are combined by interpolation with a factor $\alpha$ after the sign of $\bI$ has been flipped, which results in a QUBO matrix, \cref{eq:Q}.
		The corresponding QUBO problem, \cref{eq:qfs}, is solved either through quantum computing or classical solvers.
		The resulting binary solution vector $\bx^*$ is a bit mask that indicates the selected features.}
	\label{fig:pipeline}
\end{figure*}

Formally, \cref{eq:qfs} represents a QUBO problem.
For this reason, we call our method \emph{QUBO feature selection}, or \emph{QFS} for short.
A QUBO objective function, \cref{eq:Q}, is typically written in a quadratic form
\begin{equation}\label{eq:fsqubo}
	Q(\bx,\alpha) = \bx^\intercal \bQ(\alpha) \bx
\end{equation}
with a QUBO matrix $\bQ(\alpha)$. The elements of this matrix read
\begin{equation}\label{eq:Qij}
	Q_{ij}(\alpha) = R_{ij}-\alpha ( R_{ij} + \delta_{ij} I_i ),
\end{equation}
where $\delta_{ij} := \indicator{i=j}$ denotes the Kronecker delta.
The solution of this QUBO instance, \cref{eq:qfs}, represents the optimal feature subset.
We provide a short review of QUBOs and their solution strategies in \cref{sec:qubo}.
A complete pipeline of how FS is performed according to our proposed framework is shown in \cref{fig:pipeline}.

\subsection{Controlling the number of selected features}
For FS, it is of particular interest to be able to select a specific number of features.
Formally, this can be realized by adding a constraint to \cref{eq:qfs} such that the selection of $k$ features is enforced.
The resulting constrained optimization problem reads
\begin{equation}\label{eq:qfs:k}
	\bx^* := \underset{\substack{x_i\in\lbrace 0,1\rbrace\, \forall i\in [n] \\ ~\subjectto ~\norm{\bx}_1=k}}{\arg\min} Q(\bx,\alpha)
\end{equation}
and is formally not a QUBO in contrast to \cref{eq:qfs}.
To come back to QUBO form, a straightforward approach is to add a penalty term such as $\lambda\left(\left(\sum_ix_i\right)-k\right)^2$ to $Q(\bx,\alpha)$, which is only equal to zero for a selection of exactly $k$ features.
Here, $\lambda$ represents a strictly positive factor called Lagrangian.
The problem can then be solved to obtain the desired solution such that $k = \norm{\bx^*}_1$.\par
However, two challenges arise from this approach.
First, a suitable choice of $\lambda$ is not clear and depends on the magnitude of $Q(\bx,\alpha)$.
If it is chosen too small, the imposed constraint might be ignored for certain solutions.
On the other hand, if it is chosen too large, it may lead to a very large value range and, consequently, to loss of precision.
Summarized, having both very small and very large elements in the QUBO matrix limits the amount of scaling to amplify meaningful differences between loss values.
Furthermore, it is not clear in the fist place whether a feasible solution to the constrained problem can be found at all.\par

Due to these difficulties, we propose an alternative strategy to specify the number of selected features.
Instead of resorting to penalty terms, we use the fact that the choice of $\alpha$ itself can be used the number of features present in the solution.
This presumption can be motivated by considering the extremal values of $\alpha \in [0,1]$.
If we set $\alpha=0$, all diagonal entries of $\bQ(0)$ become $0$ and we put full emphasis on redundancy.
Trivially, both the empty set of features as well as any single feature is least redundant, so that the optimally selected set of features is either $\lbrace\emptyset\rbrace$ or $\lbrace\lbrace i\rbrace\,\vert\, i\in [n]\rbrace$.
Conversely, if $\alpha=1$, the problem becomes linear with only negative coefficients, leading to a selection of all features as the optimal solution.
Consequently, by iteratively varying $\alpha$ from $0$ to $1$ in sufficiently small steps, we observe that $\lVert\bx^*\rVert_1$ increases monotonically in steps of one from $0$ to $n$.
This observation suggests that we can tune $\alpha$ to obtain a subset of any desired size $k\in\lbrace 0,1,\dots,n\rbrace$.
As we show with \cref{prop:alpha}, this assumption is indeed true.

\begin{proposition}\label{prop:alpha}
	For all $Q(\cdot, \alpha)$ defined as in \cref{eq:fsqubo} and $k\in\lbrace 0,1,\dots,n\rbrace$, there is an $\alpha\in [0,1]$ such that $\bx^*\in\arg\min_{\bx}Q(\bx,\alpha)$ and $\norm{\bx^*}_1=k$.
\end{proposition}

The proof can be found in \cref{sec:proof}.
This result shows that we do not need additional constraints on the QUBO instance to control the number of features present in the global optimum.

\subsection{QFS Algorithm}
\label{sec:qfsalg}

From the result of \cref{prop:alpha}, we can devise an algorithm that, if provided with $\bR$, $\bI$ and $k$, returns an $\alpha^*$ such that an optimal feature subset vector $\bx\in Q^*(\alpha^*)$ has exactly $k$ non-zero entries.
For this purpose, we introduce
\begin{equation}\label{eq:Q:k}
	Q^*_k(\alpha) := \underset{\substack{\bx\in\lbrace 0,1\rbrace^n \\ ~\subjectto ~\norm{\bx}_1=k}}{\min} Q(\bx,\alpha)
\end{equation}
with $0\leq k\leq n$, \ie, the minimal function value of $Q(\bx,\alpha)$, \cref{eq:fsqubo}, for a given $\alpha$ when the number of ones in the solution is restricted to $k$. Furthermore, we denote the minimum with respect to $k$ (\ie, the global minimum) by
\begin{equation}\label{eq:Q:k:min}
	Q^*(\alpha) := \min_k Q^*_k(\alpha).
\end{equation}
Assuming we have an oracle for $Q^*(\alpha)$ that, given any $Q(\cdot,\alpha)$, returns a global optimum, an appropriate value for $\alpha^*$ can be found in $\cO(\log n)$ steps through binary search.
The value of $\alpha^*$ is not necessarily unique.

In addition to the procedure described above, we introduce a threshold $\epsilon\geq 0$, such that $Q_{ii}(\alpha)$, \cref{eq:Qij}, is set to some small positive value $\mu>0$ if $\alpha I_i<\epsilon$ for all $i\in [n]$. That is, we perform the transition $Q_{ij}(\alpha) \mapsto Q_{ij}(\alpha,\epsilon, \mu)$ with
\begin{equation}\label{eq:Qij:epsilon}
	Q_{ij}(\alpha,\epsilon, \mu) = \begin{cases} \mu & \text{if $i=j \,\land\, \alpha I_i<\epsilon$} \\ Q_{ij}(\alpha) & \text{otherwise} \end{cases}
\end{equation}
for arbitrary but constant values of $\epsilon$ and $\mu$.
In our experiments we observed that, when the importance value of feature $i$ is very close to zero, it has virtually no influence on the function value, and the solvers we used tended to include or exclude it randomly.
As we seek to minimize the number of necessary features, we add an artificial weight $\mu$ to exclude these features from the optimal solution and avoid randomness.
The exact value of $\mu$ is not decisive as long as it is positive, which ensures that the respective feature cannot be part of an optimal solution \cite[Lemma 1.0]{glover_logical_2018}.

The proposed algorithm is sketched in \cref{alg:alpha}.
This algorithm is the main contribution of this manuscript.

\begin{algorithm}
	\KwData{$\bR$, $\bI$, $\epsilon$, $\mu$, $k$}
	\KwResult{$\alpha^*$ and $\bx^*$ with $\lVert\bx^*\rVert_1=k$}
	$\:\!$ $a\gets 0$\\
	$b\gets 1$\\
	$\alpha\gets 0.5$\\
	$\bx^*\gets Q^*(\alpha,\epsilon,\mu)$\\
	$k'\gets \norm{\bx^*}_1$\\
	\While{$k'\neq k$}{
		\eIf{$k'>k$}{
			$b\gets \alpha$\\
		}{
			$a\gets \alpha$\\
		}
		$\:\!$ $\alpha\gets (a+b)/2$\\
		$\bx^*\gets Q^*(\alpha,\epsilon,\mu)$\\
		$k'\gets \norm{\bx^*}_1$\\
	}
	\Return{$\alpha$, $\bx^*$}
	\caption{Our proposed QFS Algorithm: Binary search that, given an integer $k$, finds a value $\alpha^*$, such that the optimal solution of a feature selection QUBO problem with matrix elements $Q_{ij}(\alpha^*, \epsilon, \mu)$, \cref{eq:Qij:epsilon}, contains $k$ features.}
	\label{alg:alpha}
\end{algorithm}

\section{Solving QUBOs}
\label{sec:qubo}

An integral part of our proposed QFS algorithm, \cref{alg:alpha}, is the solution of QUBOs, \cref{eq:qfs}.
QUBOs of the form
\begin{equation}\label{eq:qubo}
	\underset{\bx\in\lbrace 0,1\rbrace^n}{\min} \bx^\intercal \bQ \bx
\end{equation}
with symmetrical $\bQ \in \mathbb{R}^{n \times n}$ as in \cref{eq:fsqubo} are a popular class of optimization problems, known to be NP-hard \citep{pardalos_complexity_1992}.
Numerous practical optimization problems have been embedded into QUBO form, ranging from finance and economics \citep{laughhunn_quadratic_1970,hammer_applications_1971} over satisfiability \citep{kochenberger_using_2005} to ML tasks such as clustering \citep{kumar_quantum_2018,bauckhage_adiabatic_nodate}, vector quantization \citep{bauckhage_hopfield_2020}, support vector machines \citep{mucke_learning_2019,date_qubo_2020} and probabilistic graphical models \citep{mucke_learning_2019}, to name a few.

\subsection{Classical solvers}
Motivated by its relevance for practical problems, a wide range of classical methods to solve QUBOs have been developed, both exactly and approximately.
A comprehensive overview of applications and solution methods can be found, \eg, in \cite{kochenberger_unconstrained_2014}.
Notable among the heuristic approaches are evolutionary algorithms and simulated annealing, for which highly efficient special-purpose hardware has been developed \citep{mucke_learning_2019,matsubara_digital_2020} that can quickly find good approximate solutions to QUBO instances with several thousand variables.
Another heuristic solver is \emph{qbsolv} \citep{booth_partitioning_2017}, which finds approximate solutions of a QUBO instance by iteratively partitioning it into smaller sub-problems, which are solved through a variation of tabu search \citep{glover_tabu_1998}.
The sub-problems are created by ``clamping'' certain bits of a current solution, \ie, treating them as fixed and only optimizing over the remaining bits.
The bits to be optimized are selected according to their impact on the objective function value, \ie, its increase when negating them in the current best solution.\par

\subsection{Quantum solvers}
\label{sec:quantumsolvers}
In recent years, quantum computing has opened up a promising approach to solving QUBO instances, which -- among its many other applications -- makes it interesting for performing FS.
For this purpose, any QUBO problem can be encoded in form of a Hamiltonian
\begin{equation}\label{eq:ising}
	\hat{H}=\sum_{\substack{i,j=0\\i \neq j}}^n a_{ij}\hat{\sigma}_i\hat{\sigma}_j + \sum_{i=0}^n b_i\hat{\sigma}_i + c,
\end{equation}
where $\hat{\sigma}_i$ denotes a Pauli-$z$ matrix acting on qubit $i$ with eigenvalues $\pm1$ and corresponding eigenstates $\ket{\pm_i}$.
The coefficients $a_{ij},b_i,c\in\mathbb{R}$ can be found by performing the transformation $x_i \mapsto (1-\hat{\sigma}_i)/2$ of the objective function in \cref{eq:qubo} with $\hat{\sigma}_i^2=1$. Since the Pauli spin matrices of different qubits commute, the minimum eigenstate of the Hamiltonian $\ket{\Psi}$ with $\hat{H} \ket{\Psi} = E \ket{\Psi}$ can be written in terms of $\ket{\Psi} = \ket{\psi_1} \otimes \cdots \otimes \ket{\psi_n}$, where $\ket{\psi_i} \in \{ \ket{+_i}, \ket{-_i} \} \,\forall\,i \in [1,n]$.
Therefore, the eigenvalue $E = \min_{\bx\in\lbrace 0,1\rbrace^n} \bx^\intercal \bQ \bx$ represents the minimum objective function value of the QUBO.
The corresponding solution vector $\bx^* = \arg\min_{\bx\in\lbrace 0,1\rbrace^n} \bx^\intercal \bQ \bx $ can be obtained from the eigenstate $\ket{\Psi}$ with the assignment $x_i^* = \vert\braket{-_i|\Psi}\vert^2$ based on a projective measurement of each qubit (and is not necessarily unique).
Summarized, the QUBO problem can be transformed into the problem of finding the minimum eigenstate (or ground state) $\ket{\Psi}$ of $\hat{H}$.

However, finding the ground state of a Hamiltonian is in general also a non-trivial problem, and possible solution strategies depend on the properties of the quantum hardware and the shape of $\hat{H}$.
Two common approaches are the Variational Quantum Eigensolver (VQE) \citep{peruzzo_variational_2014}, which is suitable for quantum gate computers, and Quantum Annealing (QA) \citep{kadowaki_quantum_1998,morita2008mathematical}, which is suitable for quantum annealers.
VQE uses a hybrid quantum-classical computational approach \citep{mcclean_theory_2016} to minimize the expected energy $\bra{\Psi(\btheta)}\hat{H}\ket{\Psi(\btheta)}$.
For this purpose, a parametric ansatz $\hat{U}(\btheta)$ is prepared on a quantum gate computer such that $\ket{\Psi(\btheta)} = \hat{U}(\btheta)\ket{0}$.
The circuit parameters $\btheta$ are learned with a classical optimization in order to find an estimate for the ground state $\ket{\Psi} \approx \ket{\Psi(\btheta)}$.
In contrast, QA exploits the adiabatic theorem \citep{farhi_quantum_2000} by preparing the ground state $\ket{\Psi_0}$ of a simple mixing Hamiltonian $\hat{H}_0$ and then slowly transferring the system into the ground state $\ket{\Psi}$ of the target Hamiltonian $\hat{H}$ through adiabatic time evolution \citep{gruber1999thermodynamics}.
A hybrid approach can be realized by splitting the initial QUBO into smaller sub-problems and solving each with QA, \eg, by using the \emph{qbsolv} strategy explained above.

Since both algorithms are heuristic, typically multiple samples (or ``shots'') of the solution are obtained from repeated measurements.

\section{Related Research}
\label{sec:related}

There are numerous approaches of defining optimal feature subsets and finding or approximating them \citep{chandrashekar_survey_2014}.

Wrapper methods directly use the performance of classification or regression models as a criterion for selecting features \citep{john_irrelevant_1994}.
As the model must be re-trained on every candidate subset, this method is very resource-intensive.
Finding the optimal subset requires brute-forcing all $2^n$ possible subsets, which is generally intractable for large feature sets and non-trivial models.
Instead, heuristic optimization schemes are often used, such as greedy search or evolutionary algorithms \citep{leardi_genetic_1992,siedlecki_note_1993}.

Filter methods use a measure of relevance, \eg, correlation or MI with the label or target variable, for ranking features and discarding those below a certain threshold.
While those methods are very easy to compute and work well in practice, they often do not consider redundancy between selected features, leading to subsets that could potentially be much smaller.
When considering redundancy, such as pairwise MI between selected features, as an additional criterion to be minimized, the optimization problem becomes non-linear and NP-hard.

In \cite{rodriguez-lujan_quadratic_nodate} this problem is formulated as a quadratic programming (QP) task, which is solved approximately by means of dimension reduction.
The resulting solution is a real-valued weight vector that is used for ranking the features.
This last step is purely heuristic, as the QP task is merely a relaxation of the corresponding QUBO problem with binary weight variables which we discuss in this article.
QUBOs become prospectively tractable through quantum computation, which is why we do not resort to approximations or relaxations in our approach to make our method feasible.

QUBO formulation based on redundancy and importance appeared recently in literature \citep{tanahashi_global_2018,otgonbaatar2021quantum}, however, the authors rely on a penalty term with Lagrange multiplier $\lambda$ for restricting the solution to $k$ features.
The choice of $\lambda$ is not trivial and can drastically increase the dynamic range of the QUBO coefficients if chosen too large, as discussed above.
In our approach, we observe that we can weigh redundancy and importance against each other in order to control the number of features present in the optimal solution, rendering any additional constraints obsolete.

Another approach relies on quantum gate computing to apply Grover's algorithm to an oracle that yields the improvement in accuracy by adding or removing single features \citep{he_quantum-enhanced_2018}.
This is merely a quantum version of a simple sequential wrapper approach, which improves the theoretical runtime for greedily selecting the next feature to be inserted or removed, as Grover's algorithm finds the minimum or maximum of a function with logarithmic time complexity.
The solution quality is the same as for a classical greedy algorithm -- or potentially worse, as Grover's algorithm is probabilistic.

\section{Experiments}
\label{sec:experiments}

In \cref{sec:method}, we have presented our novel QFS algorithm, \cref{alg:alpha}. The current section contains a study of four different experiments to evaluate the performance of QFS.

For this purpose, we use six data sets, both synthetic and taken from real-world data sources. All data sets are listed in \cref{tab:datasets}, a detailed description can be found in \cref{sec:datasets}. For discretizing the data as described in \cref{sec:qfs}, we choose $B=20$. Furthermore, we set $\epsilon = 10^{-8}$ and $\mu=\max_{i,j\in[n]}Q_{ij}(\alpha)$ in \cref{alg:alpha}. These parameters are determined empirically by careful testing.

\begin{table}[t]
	\caption{Data sets used for our numerical experiments. For each data set, we list the number of features $n$, the number of classes $c$ and the number of samples $S$. A detailed description can be found in \cref{sec:datasets}.}
	\label{tab:datasets}
	\centering
	\begin{tabular}{llccc}
		\hline\noalign{\smallskip}
		Name & References & $n$ & $c$ & $S$ \\
		\noalign{\smallskip}\hline\noalign{\smallskip}
		\texttt{mnist} & \cite{lecun_mnist_2010} & \num{784} & \num{10} & \num{70000} \\
		\texttt{ionosphere} &\cite{sigillito_classification_1989,dua_uci_2017} &\num{34} &\num{2} &\num{351} \\
		\texttt{waveform} & \cite{breiman_classification_1984,dua_uci_2017} & \num{21} & \num{3} & \num{5000} \\
		\texttt{madelon} & \cite{guyon_result_2004,dua_uci_2017} & \num{500} & \num{2} & \num{2000} \\
		\texttt{synth\_10} & & \num{10} & \num{2} & \num{10000} \\
		\texttt{synth\_100} & & \num{100} & \num{2} & \num{10000} \\
		\noalign{\smallskip}\hline
	\end{tabular}
\end{table}

Our first experiment in \cref{sec:mnistsubsets} serves as a proof of principle, in which we evaluate whether QFS is able to find any useful features at all. This is realized by selecting 30 features from the \texttt{mnist} data set through QFS and training separate 1-vs-all classifiers (on all digits). To evaluate the usefulness of the selected features, we compare the performance of these classifiers to those trained on 30 random features and all available features, respectively.
The second experiment in \cref{sec:othermethods} is a wider empirical comparison of various combinations of commonly used FS methods and ML models with the goal to show that QFS is competitive.
In \cref{sec:autoencoder}, we apply QFS in a more application-oriented setting by using the selected features as a means of lossy data compression, interpreting the reduced feature space as a latent representation and training a convolutional neural network to reconstruct the original features.
Finally, we use quantum hardware in \cref{sec:vqe} to solve two exemplary feature selection QUBO instances for QFS. This experiment demonstrates that our method can indeed be used with current NISQ devices.

\subsection{Experiment 1: Feature Quality}
\label{sec:mnistsubsets}

In the first experiment serves as a proof of principle, in which we verify that QFS is able to find informative features.

\subsubsection{Setup}
For this purpose, we consider the \texttt{mnist} data set and run \cref{alg:alpha} ten times with $k=30$ such that we use about \SI{3.8}{\percent} of the original \num{784} features (\ie, pixels).
To solve the QUBO instances classically, we use \emph{qbsolv} \citep{booth_partitioning_2017}.
As redundancy, we use the pairwise MI matrix over all features, \cref{eq:Rij}.
As importance for digit $d$, we calculate the MI between the features $x_i$ and a binary variable $\indicator{y=d}$, \cref{eq:Ii}.
This yields ten feature selections, each tailored towards one digit, which we use to perform a 1-vs-all classification.
To quantify whether the selected features from our method are informative or not, we train a Random Forest classifier on these features and determine its accuracy.
For comparison, we also determine the accuracy of a Random Forest classifier that has been trained on the whole feature set and a set of randomly selected features (uniformly sampled from the set of $k$-element subsets of $[n]$ for each digit), respectively.
Specifically, the Random Forest is composed of 100 estimators, each a Decision Tree of maximal depth five and a maximum of five features considered when searching the best split. These restrictions serve to limit the model size, as is a common objective in applications which require FS.
We use the Python implementation provided by \emph{scikit-learn} \citep{pedregosa_scikit-learn_2011}.

\subsubsection{Results}
\begin{figure*}\label{fig:mnistfeatures}
	\centering
	\includegraphics[width=\textwidth]{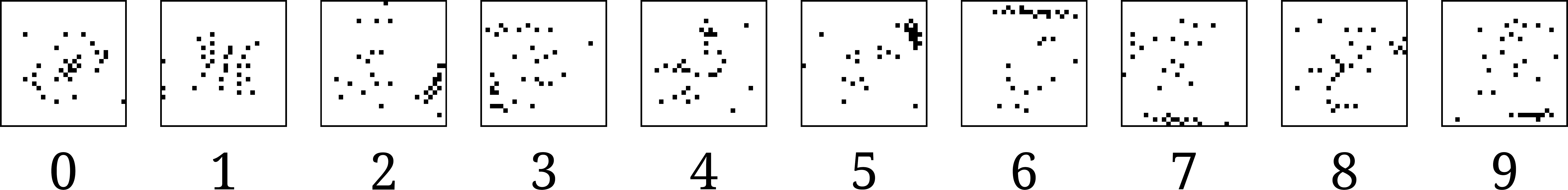}
	\vspace{1mm}
	\caption{Experiment 1: Feature subsets found through QFS on all separate digits of the \texttt{mnist} data set.
		The black pixels represent the selected features.
		The digits are ordered from left to right, starting with 0 on the left.}
\end{figure*}

\begin{figure*}
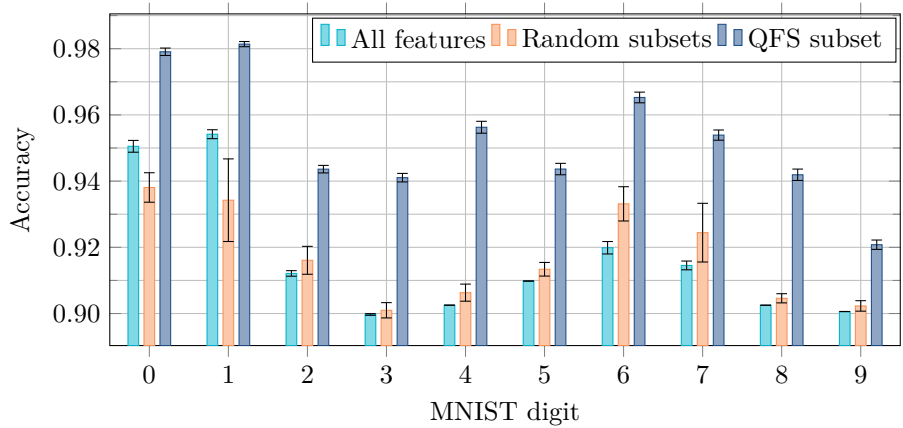

	\centering
	\includestandalone{fig_experiment4}
	\caption{Experiment 1: 10-fold cross-validated accuracy of binary Random Forest classifiers of each separate digit of the \texttt{mnist} data set. The classifiers are trained on (1) all \num{784} features, (2) 30 randomly sampled features, and (3) 30 features selected by QFS.
		The accuracy (10-fold CV) using random features is averaged over 5 different random subsets.
		The standard deviation (one sigma) is represented by the smaller error bars on top of the bars, which show the mean.}
	\label{fig:exp4}
\end{figure*}

The selected features (pixels) per digit are shown in \cref{fig:mnistfeatures}.
We perform 10-fold cross validation and report mean and standard deviation of the classification accuracy, which is visualized in \cref{fig:exp4}.
For ``Random subsets'', we report the cross-validated accuracy averaged over five random subsets, so mean and standard deviation is reported over 50 runs in total.

We observe that the Random Forest model using the constrained estimators achieves the best results on all digits using the optimal features found through QFS.
This confirms that our method indeed finds informative features that are useful for classification.
Interestingly, the models trained on the QFS subsets not only outperform the models trained on random subsets, but also those using all features.
This is probably due to the restricted number of features per split in the base estimators, which leads to a higher chance of picking non-informative features when using all available pixel values.

\subsection{Experiment 2: Cross-Model Comparison with FS Methods}
\label{sec:othermethods}

In the second experiment, we contrast our method of QFS to other feature selection methods by comparing the accuracy of various classification models trained on the respective feature subsets in analogy to the previous experiment from \cref{sec:mnistsubsets}. 

\subsubsection{Setup}
Specifically, we apply \cref{alg:alpha} to five different data sets to obtain feature subsets of fixed sizes $k$. We consider 30 features (\SI{3.8}{\percent}) for \texttt{mnist}, five features (\SI{14.7}{\percent}) for \texttt{ionosphere}, 20 features (\SI{4}{\percent}) for \texttt{madelon}, 20 features (\SI{5}{\percent}) for \texttt{synth\_100}, and 5 features (\SI{23.8}{\percent}) for \texttt{waveform}.
For \texttt{madelon} and \texttt{synth\_100} we use the known number of informative features, while for the remaining data sets we chose the feature subset sizes arbitrarily in varying percentage ranges of the total number of features.

To evaluate the quality of the selected features, we compare QFS to three other heuristic FS methods:
\begin{enumerate}
	\item Ranking obtained from the Euclidean norm over the coefficients of $n$ 1-vs-all Logistic Regression (LR) models.
	\item Ranking obtained from the impurity-based feature importances given by an Extra Trees Classifier (ET) with \num{100} estimators.
	\item Recursive Feature Elimination (RFE) \citep{guyon_gene_2002} performed on a Decision Tree of maximal depth \num{10}.
\end{enumerate}

The resulting feature subsets are then used to train five classification models:
\begin{enumerate}
	\item Neural Network
	\item 1-vs-all Logistic Regression
	\item Decision Tree
	\item Random Forest (of 100 decision trees)
	\item Naive Bayes classifier (with Gaussian prior)
\end{enumerate}

The Neural Network has a single hidden layer containing $\lfloor\sqrt{k}+0.5\rfloor$ neurons to make the dependency between the number of parameters and selected features $k$ linear.
For the activation function we use ReLU.
Both the Neural Network and the Logistic Regression are limited to 1000 learning iterations.
Again, we use Python implementations provided by \emph{scikit-learn} \citep{pedregosa_scikit-learn_2011} for all models and FS methods.

On every model we perform a 10-fold cross validation and report mean and standard deviation of the classification accuracy.

\subsubsection{Results}

\begin{figure*}
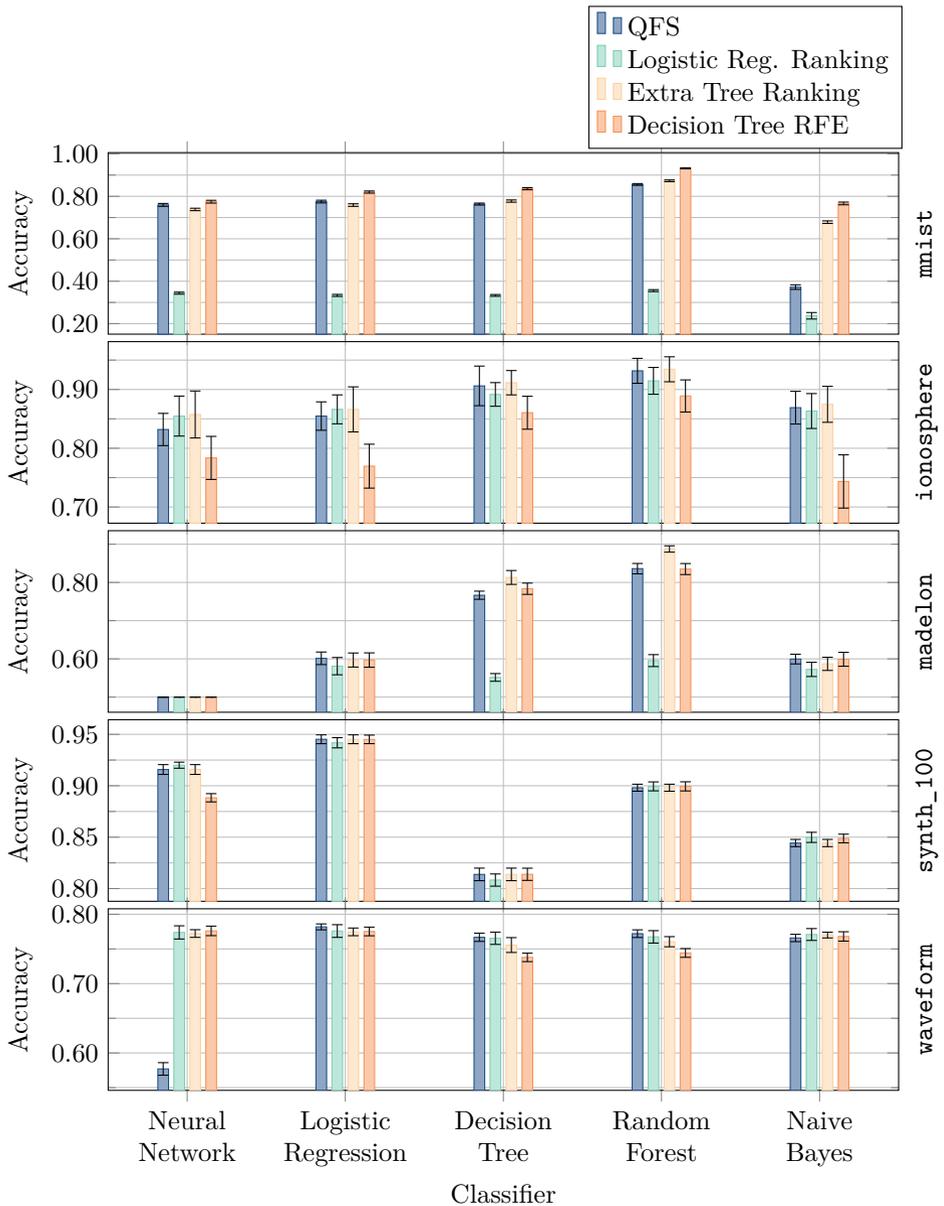

	\centering
	\includestandalone{fig_experiment6}
	\caption{Experiment 2: 10-fold cross-validated accuracy of various classifiers using feature subsets produced by various feature selection methods on three different data sets. The standard deviation (one sigma) is represented by the smaller bars on top of the bars showing the mean.}
	\label{fig:exp6}
\end{figure*}

The results are visualized in \cref{fig:exp6}.
The results of this experiment show that QFS, in general, compares favorably among FS methods.
The RFE method using decision trees often leads to better accuracies, but comes at the cost of much higher computation time, owing to the fact that it is a wrapper method which requires the model to be re-trained in every iteration.

\begin{figure*}
	\centering
	\begin{subfigure}[t]{0.65\textwidth}
		\centering
		\includestandalone[height=.8\textwidth]{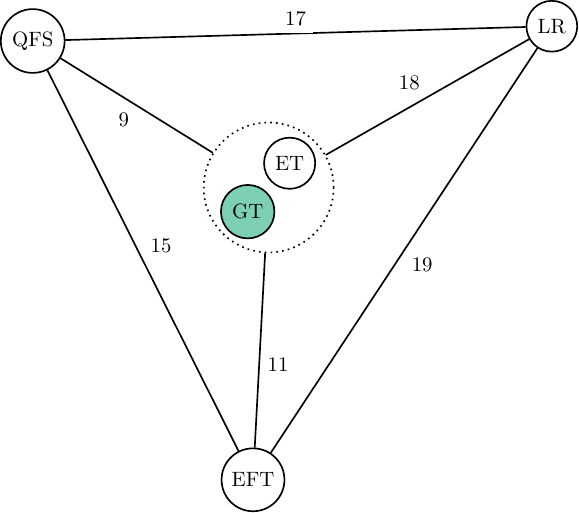}
		\caption{\texttt{madelon}}
	\end{subfigure}%
	~ 
	\begin{subfigure}[t]{0.35\textwidth}
		\centering
		\includestandalone[height=.8\textwidth]{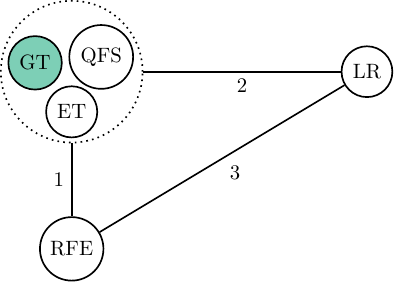}
		\caption{\texttt{synth\_100}}
	\end{subfigure}
	\caption{Experiment 2: Edit distances between feature subsets found through the different feature selection methods on the data sets \texttt{madelon} and \texttt{synth\_100} for which the ground truth informative features are known.
		Each node represents a FS method, each edge weight is the edit distance, \ie, the number of features that need to be swapped in order to convert one selection into the other.
		We show the ground truth (GT) as well as Logistic Regression ranking (LR), Extra Tree classifier ranking (ET), Recursive Feature Elimination (RFE) and our method (QFS).
		Methods within the dotted circles yield the same feature subset.}
	\label{fig:editdist}
\end{figure*}

For the data sets \texttt{madelon} and \texttt{synth\_100} we know the ground-truth informative features, which allows us to evaluate the distance between the optimal feature subset and the subset found by each method.
To this end, we use the edit distance between pairs of feature subsets, which is the number of features that need to be swapped in order to turn one subset into the other.
Alternatively, it is the Hamming distance between the binary feature indicator vectors, divided by two.
We represent each FS method used in this experiment as a node in an undirected graph and, and each edit distance between the feature vectors they produced as a weighted edge.
Nodes of distance 0 are represented as cluster nodes.
The resulting graphs are shown in \cref{fig:editdist}.

QFS is able to find all informative features for \texttt{synth\_100}, and is closest to ground-truth on \texttt{madelon} among all other methods except for the Extra Tree classifier ranking, which is able to find the ground-truth features in both cases.
This result shows that MI, when used as measure of redundancy and importance in QFS, produces useful feature subsets on both subsets where ground truth is known.
Moreover, FS on these data sets seems to work very well with the Extra Tree classifier ranking method, which indicates that impurity-based measures are particularly effective here.

\subsection{Experiment 3: Application: Lossy Compression with Autoencoder}
\label{sec:autoencoder}

\begin{figure}
	\centering
	\includegraphics[width=\textwidth]{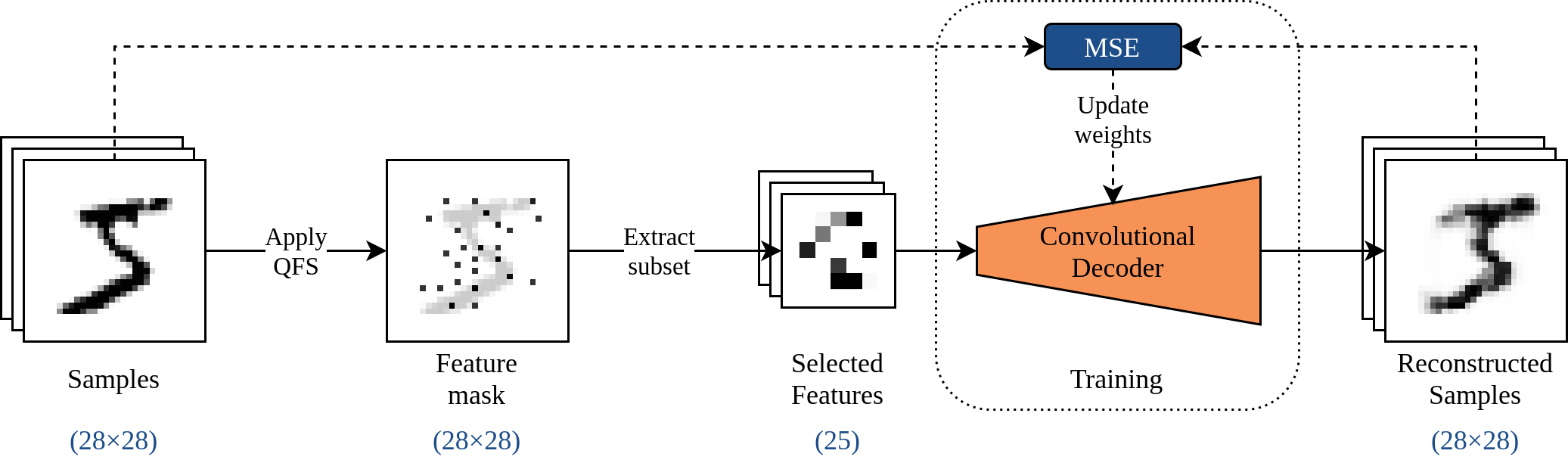}
	\caption{Experiment 3: Compression and decompression pipeline using QFS with $k=25$ and a convolutional decoder.
		The features are being extracted by applying the feature selection mask to the image, which yields the compressed representation (compression rate of $25/784=\SI{3.19}{\percent}$).
		Then the decoder is trained on minimizing the mean squared error between a reconstruction of the original image.
		The feature mask shown above is the actual feature mask used in the experiment.}
	\label{fig:ae-pipeline}
\end{figure}

The previous experiments have confirmed that QFS picks features of a data set which are important and carry little redundancy according to our criteria of choice.
From a different perspective, this can be interpreted as removing unimportant and redundant features, which leads to much smaller data sets.
Consequently, QFS can be used as a type of lossy compression by computing an optimal feature subset $S$ and discarding all features $i\notin S$.
From this compressed representation we can reconstruct the original data points approximately by means of some ML model \citep{kingma2013auto}.
This process is shown schematically in \cref{fig:ae-pipeline}.
In this third experiment, we evaluate the lossy compression empirically.

\subsubsection{Setup}
To realize this experiment, we use \cref{alg:alpha} to perform QFS on the \texttt{mnist} data set with $k=25$ (\ie, \SI{3.19}{\percent} of all features).
The resulting subset $S$ of pixel positions is interpreted as a latent space, \ie, a compressed data representation like one we would obtain from principal component analysis (PCA) or an autoencoder.
While PCA is based on an eigendecomposition that is used to project data onto axes of maximal variance, the subspace $S$ induced by the feature selection fulfills other optimality criteria based on redundancy and importance.

By feeding this latent representation into a convolutional neural network (CNN), we can reconstruct the original images by projecting back to a size of $28\times 28$ and minimizing the difference between reconstruction and original.
To this end, our CNN architecture consists of a linear input layer with $k$ inputs and \num{392} outputs.
The output is reshaped to 8 channels of size $7\times 7$.
Using two sequential 2D transposed convolution operations interspersed with ReLU activations, the images are first inflated to $16\times 14\times 14$ and finally to $1\times 28\times 28$.
A sigmoid is applied to the output in order to obtain pixel values between 0 and 1.
The model weights are trained by minimizing the mean squared error (MSE) between the original samples $\bx$ and the reconstructions,
\begin{equation}
	\frac{1}{784}\norm{\bx-f_{\btheta}(\bx_S)}_2^2 ~\rightarrow ~\min_{\btheta},
\end{equation}
where $f_{\btheta}$ is the model function with weights $\btheta$, and $\bx_S$ the sub-vector of $\bx$ containing only the features in $S$ found through QFS.
We train the model for \num{1000} epochs with batches of size \num{250}, using the Adam optimizer \citep{kingma_adam_2014} from \emph{PyTorch} \citep{paszke_pytorch_2019} with a 1cycle learning rate scheduler \citep{smith_super-convergence_2018} with maximum learning rate \num{0.01}.

\begin{figure*}[t!]
	\centering
	\begin{subfigure}[t]{0.5\textwidth}
		\centering
		\includegraphics[height=.8\textwidth]{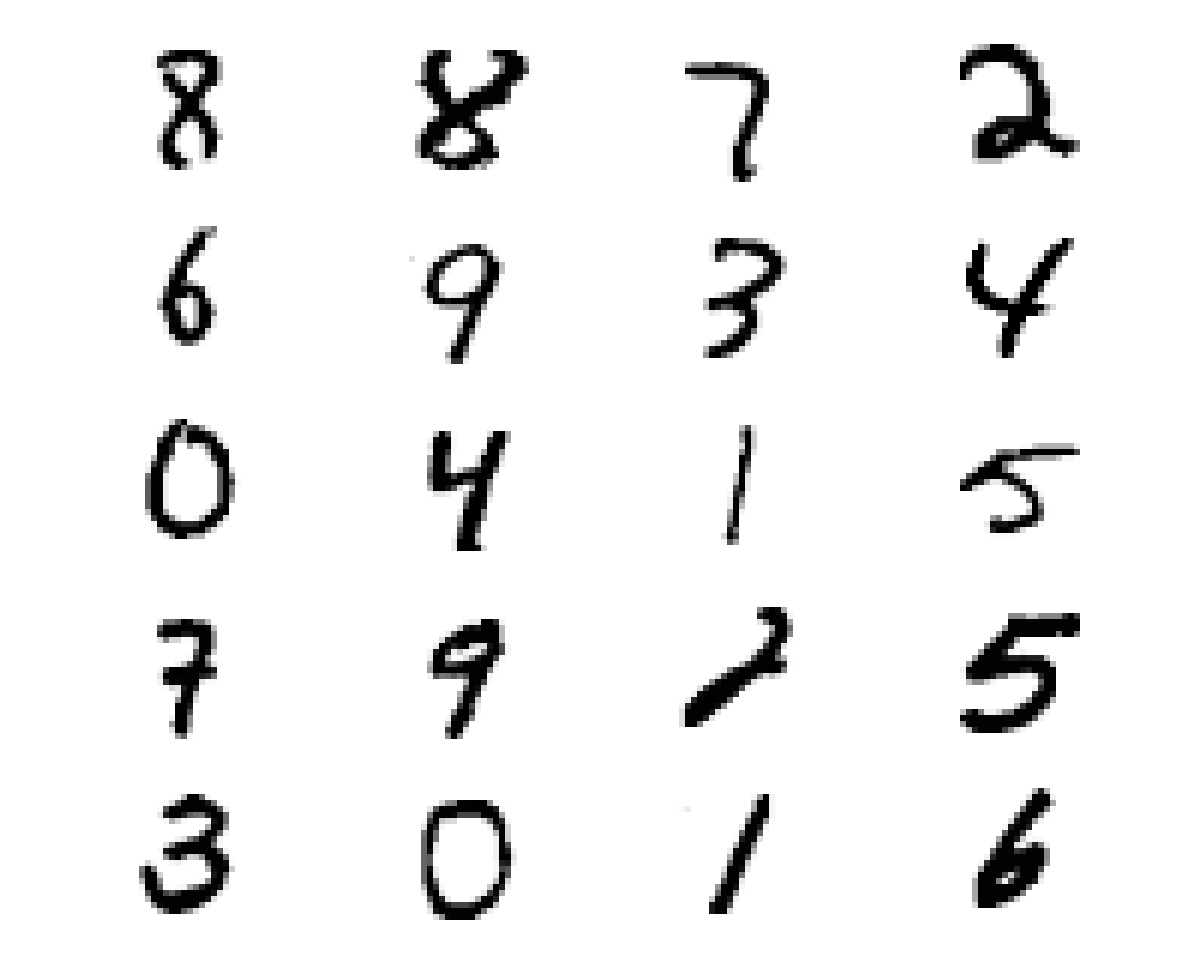}
		\caption{Original samples}
	\end{subfigure}%
	~ 
	\begin{subfigure}[t]{0.5\textwidth}
		\centering
		\includegraphics[height=.8\textwidth]{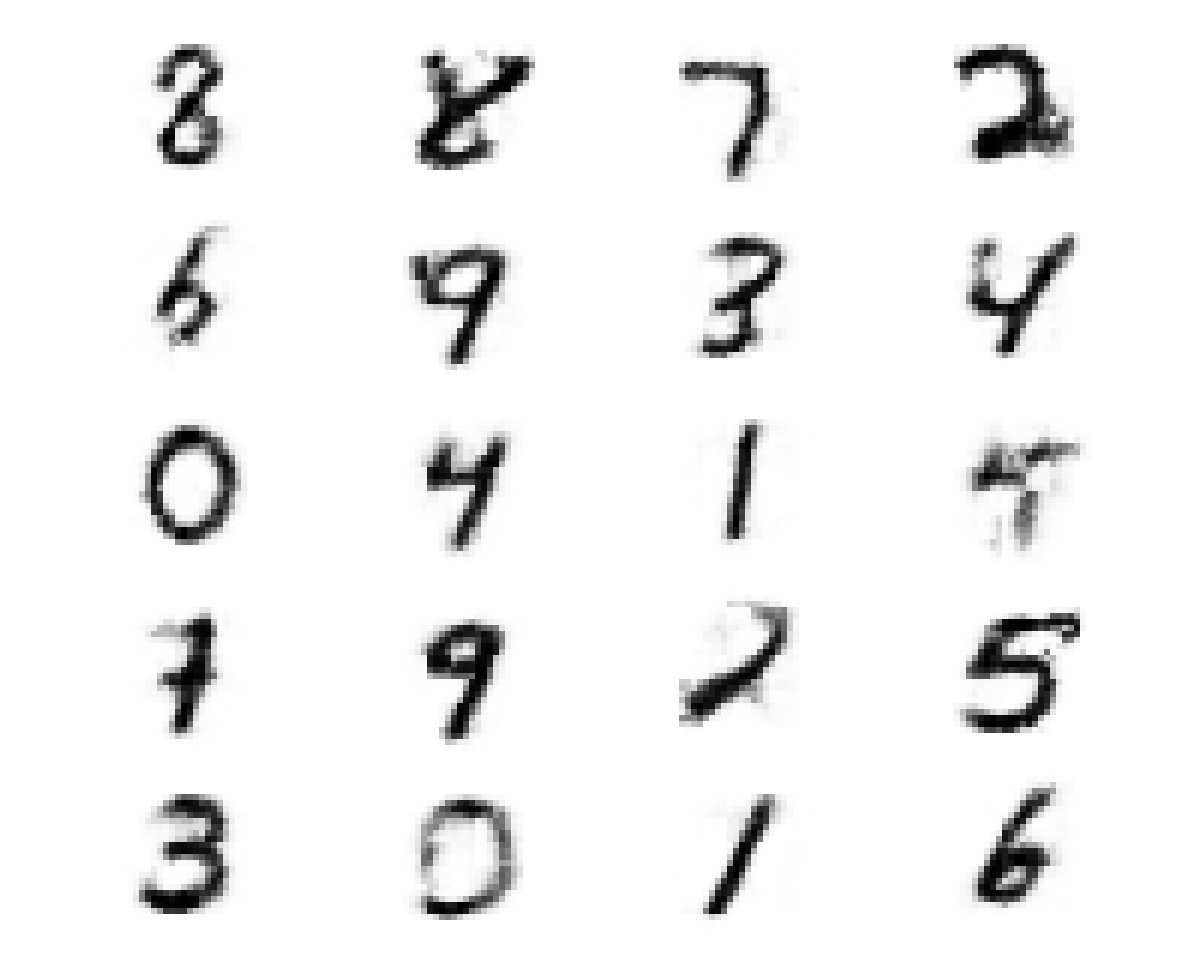}
		\caption{Reconstruction}
	\end{subfigure}
	\caption{Experiment 3: Visual comparison of \texttt{mnist} samples reconstructed from 25 pixels at fixed positions selected through QFS.
		The reconstruction is implemented by a convolutional neural network.}
	\label{fig:ae-samples}
\end{figure*}

\subsubsection{Results}
The CNN achieves a MSE of \num{21.7389}, which corresponds to an average squared pixel deviation of \num{0.0277}.
\Cref{fig:ae-samples} shows \num{20} random \texttt{mnist} samples on the left, and their respective reconstructions using the procedure described above.

The reconstructed samples are visually very similar to the originals, which suggests that the pixel positions found through QFS contain useful information about the samples that help to deduce the values of nearby pixels.

\subsection{Experiment 4: QFS on Quantum Hardware}
\label{sec:vqe}

So far, we have only used classical QUBO solvers to perform QFS.
In this final experiment, we consider QFS with actual quantum hardware as a QUBO solver, using both a quantum annealer as well as a quantum gate computer.

\subsubsection{Setup}

Based on QFS, we construct QUBO instances for three data sets, \texttt{ionosphere}, \texttt{waveform}, and \texttt{synth\_10} as before, but solve them using QA on a \emph{D-Wave} quantum annealer and VQE on an \emph{IBM} gate quantum computer as described in \cref{sec:quantumsolvers}.

To conduct this experiment, we first perform an entirely classical run of \cref{alg:alpha} to obtain values for $\alpha$ for a predefined number of selected features $k$.
In principle, this search for $\alpha$ can be done on quantum hardware as well.
However, we can reduce the quantum computation time significantly by pre-computing $\alpha$ classically.
This is crucial since access of the quantum gate hardware is time-consuming, especially for the IBM gate quantum computer. For this reason, we also only consider the \texttt{synth\_10} data set for the VQE algorithm.
The resulting classically obtained values of $\alpha$ are listed in \cref{tab:hardware-alphas} together with the chosen number of selected features $k$.
Using these values, we assemble a QUBO instance for each data set and solve these QUBO instances using both hardware approaches.

\begin{table}[t]
	\caption{Experiment 4: For each of the three data sets of interest, we list the number of features $n$, the chosen number of selected features $k$ and the resulting value of $\alpha$ from a classical evaluation of \cref{alg:alpha} used for the quantum experiment.}
	\label{tab:hardware-alphas}
	\centering
	\begin{tabular}{lccc}
		\hline\noalign{\smallskip}
		data set & $n$ & $k$ & $\alpha$ \\
		\noalign{\smallskip}\hline\noalign{\smallskip}
		\texttt{ionosphere} & \num{34} & \num{5} & \num{0.90625} \\
		\texttt{waveform}   & \num{21} & \num{5} & \num{0.78125} \\
		\texttt{synth\_10}  & \num{10} & \num{4} & \num{0.87500} \\
		\noalign{\smallskip}\hline
	\end{tabular}
\end{table}

The D-Wave quantum annealer \emph{Advantage~5.1} is accessed via D-Wave's cloud service \emph{Leap}\footnote{\url{https://cloud.dwavesys.com/leap}}.
It operates on \num{5627} qubits \citep{advantage2021}.
We use the QA implementation provided by \emph{ocean}\footnote{\url{https://docs.ocean.dwavesys.com/en/stable/index.html}} with the \emph{DWaveSampler} and default parameters.
We evaluate the QUBO instances for all three data sets.
In total, we obtain \num{1024} samples per QUBO instance, each representing an estimate of the solution.
Additionally, we perform the same experiment using Simulated Annealing, again using D-Wave's Python implementation contained in \emph{ocean} with default parameters.

The IBM quantum gate computer \emph{ibmq\_ehningen}, on the other hand, operates on a \emph{Falcon r5.11} processor with \num{27} qubits\footnote{\url{https://research.ibm.com/interactive/system-one/}}.
It is accessible via IBM's cloud service \emph{IBM Quantum}\footnote{\url{https://quantum-computing.ibm.com}}.
We use Qiskit Runtime's default VQE implementation \citep{qiskit2021} with the Simultaneous Perturbation Stochastic Approximation (SPSA) optimizer \citep{spall1998overview}.
We let the optimizer run for 32 iterations with Qiskit's default parameters.
As our parametric ansatz, we chose a Pauli Two-Design \citep{nakata_unitary_2017} with four layers.
Since the computations on the quantum gate hardware is time-consuming, we only evaluate the QUBO instances for one of the three data sets, the \texttt{synth\_10} data set.
Again, we obtain \num{1024} samples from the resulting circuit.

\subsubsection{Results}

\begin{figure}[!t]
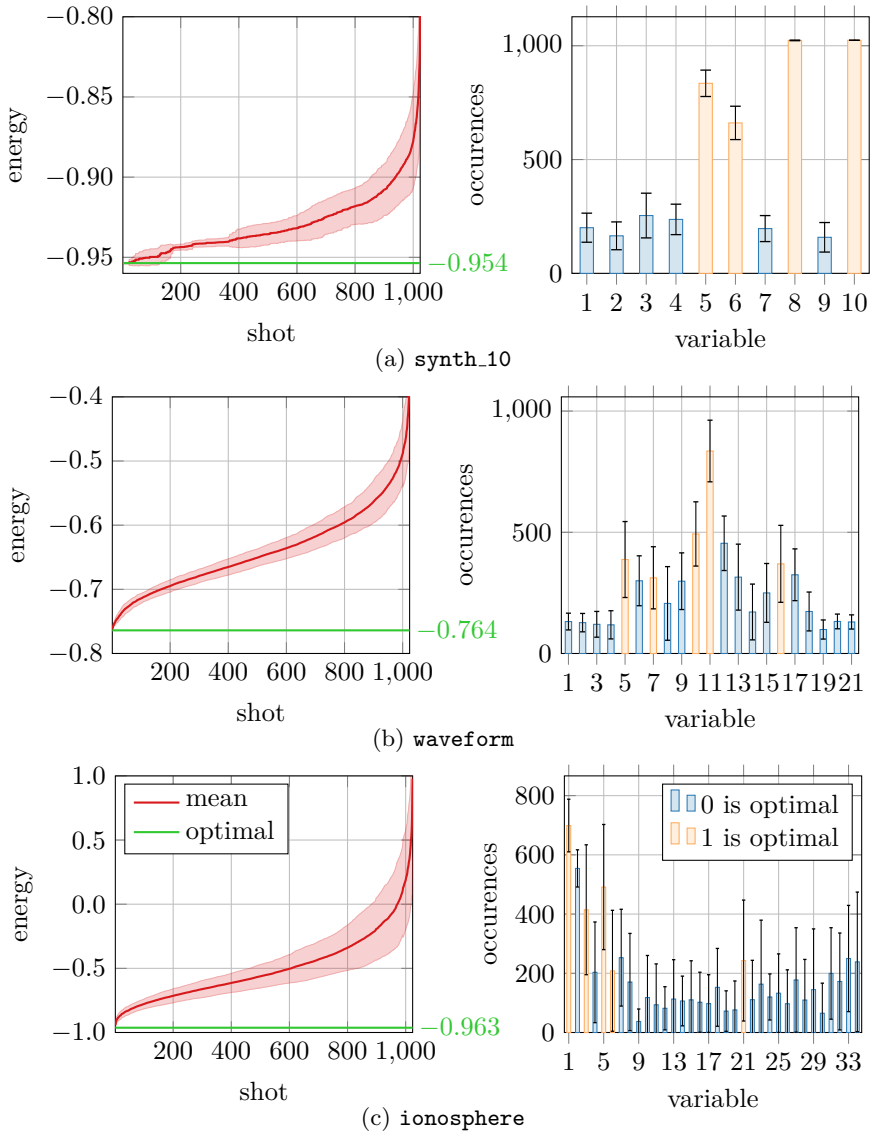

	\centering
	\begin{subfigure}[t]{1\linewidth}
		\centering
		\includestandalone{dwave21}
		\vspace{-.35cm}%
		\caption{\texttt{synth\_10}}
		\label{fig:qashots:synth}
	\end{subfigure}%
	\\\vspace{-.2cm}%
	\begin{subfigure}[t]{1\linewidth}
		\centering
		\includestandalone{dwave22}
		\vspace{-.35cm}%
		\caption{\texttt{waveform}}
		\label{fig:qashots:waveform}
	\end{subfigure}%
	\\\vspace{-.2cm}%
	\begin{subfigure}[t]{1\linewidth}
		\centering
		\includestandalone{dwave23}
		\vspace{-.35cm}%
		\caption{\texttt{ionosphere}}
		\label{fig:qashots:ionosphere}
	\end{subfigure}%
	\caption{Experiment 4: Histograms of samples from the D-Wave quantum annealer performing QFS on three different data sets.
		For each run, \num{1024} samples were performed and the energy and occurrences of all solutions recorded.
		Left: Energy values of all shots sorted in ascending order, reported as mean and standard deviation over \num{16} runs.
		The globally optimal energies (found by brute force) are shown as horizontal lines.
		Right: Index of each solution bit (corresponding to feature index for QFS) versus the number of times the corresponding bit was measured to be 1 over all shots, again reported as mean and standard deviation over all 16 runs.
		The color of each bar indicates whether the corresponding global optimum of the respective bit is 0 or 1. This global optimum represents the optimal selection of features as given by QFS.}
	\label{fig:qashots}
\end{figure}

\begin{figure}[!t]
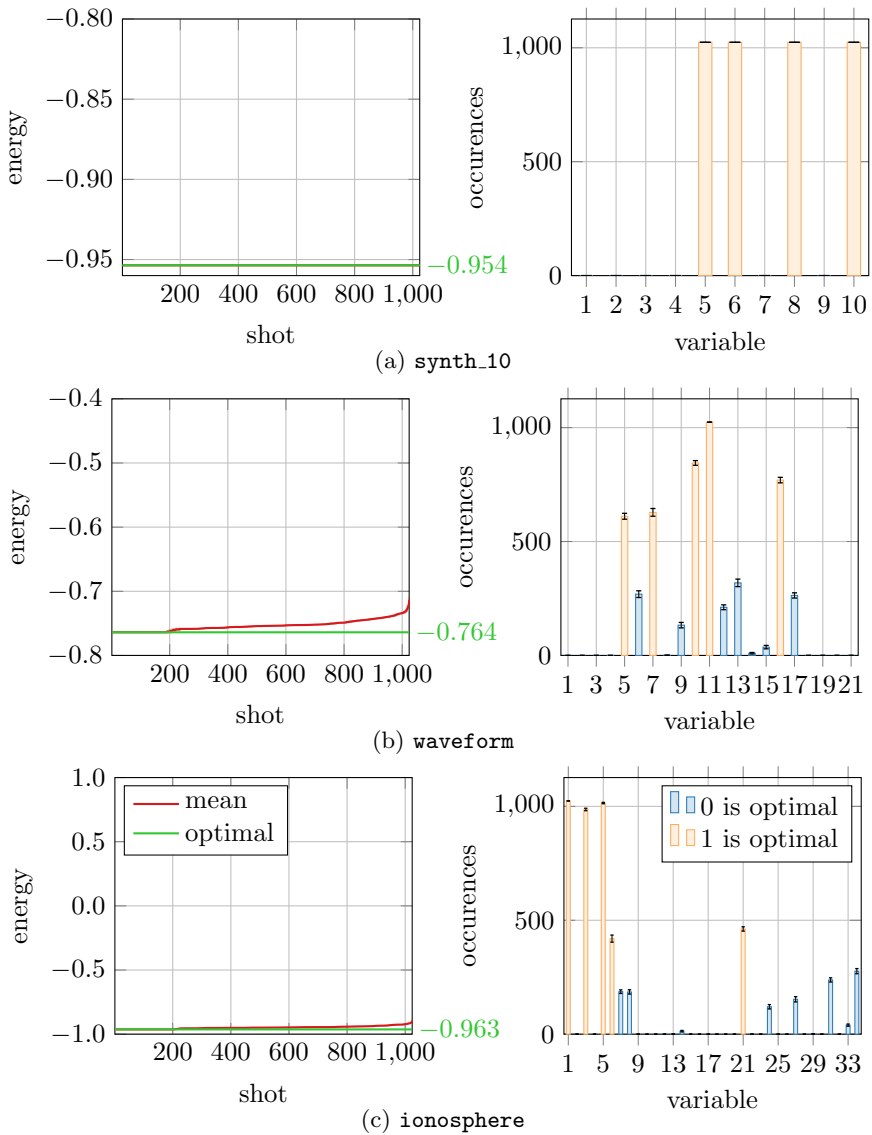

	\centering
	\begin{subfigure}[t]{1\linewidth}
		\centering
		\includestandalone{dwave21-sim}
		\vspace{-.35cm}%
		\caption{\texttt{synth\_10}}
		\label{fig:qashots:synth}
	\end{subfigure}%
	\\\vspace{-.2cm}%
	\begin{subfigure}[t]{1\linewidth}
		\centering
		\includestandalone{dwave22-sim}
		\vspace{-.35cm}%
		\caption{\texttt{waveform}}
		\label{fig:qashots:waveform}
	\end{subfigure}%
	\\\vspace{-.2cm}%
	\begin{subfigure}[t]{1\linewidth}
		\centering
		\includestandalone{dwave23-sim}
		\vspace{-.35cm}%
		\caption{\texttt{ionosphere}}
		\label{fig:qashots:ionosphere}
	\end{subfigure}%
	\caption{Experiment 4: Same results as in \cref{fig:qashots}, but using Simulated Annealing instead of the D-Wave machine.
		Optimal bits are found more frequently compared to QA, which is probably due to additional noise introduced by the quantum hardware device.}
	\label{fig:qashots-sim}
\end{figure}

The results for QA are shown in \cref{fig:qashots}.
On the left, we show all \num{1024} samples sorted in ascending order of energy, such that the lowest measured energies are on the left.
Mean and standard deviation is reported over the sorted sequences of \num{16} runs.
The globally optimal energies, which we found by brute force, are shown as horizontal lines.
On the right, we show for each solution bit (corresponding to feature indices for QFS) the number of times the corresponding bit was measured as 1 across all shots.
Again we report mean and standard deviation over \num{16} runs.
The color of each bar indicates whether the corresponding global optimum of the respective bit is 0 or 1. The sequence of optimal bits corresponds to the optimal feature selection for our application.
\Cref{fig:qashots-sim} is analogous to \cref{fig:qashots}, showing the results obtained through Simulated Annealing.

The histograms show that there is a clear correspondence between feature optimality (bar color) and the number of occurrences, which indicates that QA is able to find the global optimum in a certain fraction of samples.
With increasing number of qubits, this correlation gets noticeably less pronounced.
To be precise, the optimum was found in $\SI{10.78\pm 5.12}{\percent}$ for \texttt{synth\_10}, $\SI{0.18\pm 0.18}{\percent}$ for \texttt{waveform}, and only a single time across all 16 runs for \texttt{ionosphere}.
In contrast, Simulated Annealing finds the correct bits with higher probability, even for data of higher dimension:
The optima are found in $\SI{100.00\pm 0.00}{\percent}$ of shots for \texttt{synth\_10}, $\SI{20.39\pm 1.40}{\percent}$ for \texttt{waveform} and $\SI{21.04\pm 1.02}{\percent}$ for \texttt{ionosphere}.
This result indicates that the use of NISQ hardware compromises the solution quality, possibly due to loss of precision when loading the QUBO parameters onto the quantum annealer, or additional noise introduced by read-out errors.
%

\begin{figure}
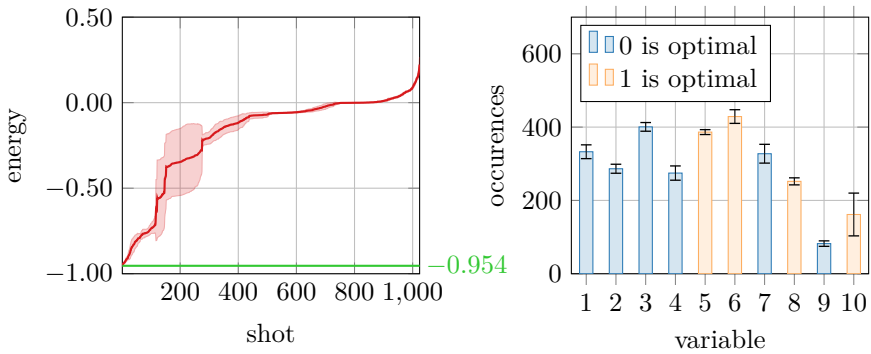

	\centering
	\includestandalone{vqe_synth10}
	\caption{Experiment 4: Histograms of samples from the IBM quantum gate computer performing QFS on the \texttt{synth\_10} data set in analogy to \cref{fig:qashots}. Measurements of all qubits from the resulting circuit are treated in the same way as the samples from the D-Wave quantum annealer. While the optimal energy (horizontal line) is reached for some samples, the overall probability to observe this global optimum is significantly lower compared with the quantum annealer. This may be due to insufficient convergence or hardware noise.}
	\label{fig:vqesynth10}
\end{figure}

The result for VQE is shown in \cref{fig:vqesynth10} in analogy to \cref{fig:qashots}.
Mean and standard deviation are obtained from \num{5} runs.
We find no clear correspondence between globally optimal bits and number of occurrences.
The global optimum was only found in $\SI{0.14\pm 0.08}{\percent}$ of measurements.
We assume that hardware noise, as well as the low number of optimization steps that were used due to long run times, lead to this performance.
In particular, we expect that VQE could perform better for longer run times and less noisy hardware.
It is important to understand that in contrast to the D-Wave results, where each shot represents one approximation run to find the underlying QFS optimum, the 1024 VQE samples were all taken from one optimized circuit.

In summary, we find that near-term quantum devices can in principle be used for QFS, especially for low data dimensions and on special-purpose hardware like quantum annealers that are designed to solve QUBO problems.

\section{Conclusion}
\label{sec:conclusion}

In this article, we present a novel algorithm for performing feature selection based on a generalized QUBO embedding, which can be solved on both classical and quantum hardware.
To this end, we use mutual information as a basis for measures of importance and redundancy, which we balance using an interpolation factor $\alpha$.
We show theoretically and empirically that our method allows the selection of feature subsets of any desired size $k$ without resorting to constraints on the solution space.

To demonstrate our framework's effectiveness, we have performed a range of experiments, comparing different common features selection methods and the resulting performance on different ML models.
Furthermore, we have also realized a practical application for lossy data compression.

One of our experiments has been run on actual quantum hardware, which further demonstrates that our algorithm is viable and NISQ-compatible.
Our experiments are conducted on rather low-dimensional problems, however, this experimental setup is dictated by the currently available hardware, and we expect our algorithm to scale in accordance with future quantum computing developments.

Our framework can easily be modified by changing measures of importance and redundancy.
Other choices instead of MI include entropy, Pearson correlation or other information-theoretic measures.
Even expert knowledge can be incorporated, if available.
As \cref{prop:alpha} is valid for general $\bI$ and $\bR$ with non-negative entries, the proof holds for any combination of importance and redundancy measures, and \cref{alg:alpha} can be applied accordingly.

\subsubsection*{Acknowledgments}
Parts of this research have been funded by the Federal Ministry of Education and Research of Germany and the state of North-Rhine Westphalia as part of the Lamarr-Institute for Machine Learning and Artificial Intelligence, LAMARR22B.
Parts of this work have been funded by the Ministry of Science and Health of the State of Rhineland-Palatinate (Germany) as part of the project AnQuC-3.
This work was developed with help of the Fraunhofer Cluster of Excellence \enquote{Cognitive Internet Technologies}.

\subsubsection*{Conflicts of Interest}
All authors declare that they have no conflicts of interest.

\subsubsection*{Data Availability}
The data that support the findings of this study are available in \url{https://github.com/Castle-Machine-Learning/feature-selection-data}.

\FloatBarrier

\newpage
\appendix

\section{Proof}
\label{sec:proof}

\begin{figure}
	\centering
	\includegraphics[width=\textwidth]{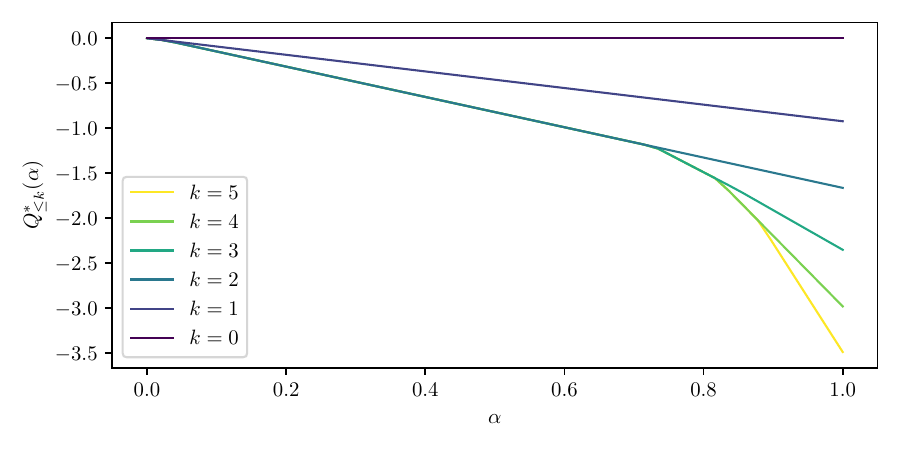}
	\caption{Graphs of $Q^*_{\leq k}$ for $k\in\lbrace 0,\dots,n\rbrace$ with $n=5$.
		Redundancy and importance values were randomly sampled.}
	\label{fig:qk}
\end{figure}

Here, we provide a proof of \cref{prop:alpha}, which states that for all $Q(\cdot,\alpha)$ defined as in \cref{eq:fsqubo} and $k\in\lbrace 0,\dots,n\rbrace$, there is an $\alpha\in [0,1]$ such that $\exists \bx^*\in\mathbb{B}^n: ~\bx^*\in\arg\min_{\bx}Q(\bx,\alpha)$ and $\norm{\bx}_1=k$.

\begin{proof}
	Recall that $R(\bx)\geq 0$ and $I(\bx)\geq 0$ for all $\bx\in\mathbb{B}^n$ due to non-negativity of mutual information.
	Firstly, note that for $\alpha=0$ both the zero vector $\bm{0}$ and all unit vectors $\bm{e}_i$ are optimal w.r.t. $Q(\cdot, 0)$, as $Q(\bm{0},0)=0$ and $\forall i\in [n]: ~Q(\bm{e}_i,0)=R_{ii}=0$ by definition of $\bm{R}$.
	This covers the cases $k=0$ and $k=1$.
	Further, if $\alpha=1$ we have $Q(\bx,1)=-I(\bx)=-\sum_{i\in [n]}I_ix_i$, which is trivially minimized by the one vector $\bm{1}$, covering the case $k=n$.
	Now, for all $k\in\lbrace 0,\dots,n\rbrace$, consider the functions \begin{equation}
		Q^*_{\leq k}(\alpha) := \min_{\bx\in\mathbb{B}^n}Q_{\alpha}(\bx)~\text{s.t. } \norm{\bx}_1\leq k\;.
	\end{equation}
	
	These functions are piece-wise linear and strictly decreasing in $\alpha$ (see \Cref{fig:qk}), due to $\partial Q_{\alpha}(\bx)/\partial\alpha=-(R(\bx)+I(\bx))\leq 0$ and non-negativity of $R(\bx)$ and $I(\bx)$.
	This implies further for all $\alpha\in [0,1]$ and $k\in [n]$ that \begin{align}
		\min_{\substack{\bx\in\mathbb{B}^n \\ \norm{\bx}_1\leq k-1}} R(x) \leq \min_{\substack{\bx\in\mathbb{B}^n \\ \norm{\bx}_1\leq k}} R(x) \\
		\max_{\substack{\bx\in\mathbb{B}^n \\ \norm{\bx}_1\leq k-1}} I(x) \leq \max_{\substack{\bx\in\mathbb{B}^n \\ \norm{\bx}_1\leq k}} I(x)\;.
	\end{align}
	This immediately implies that for any $k<k'$ \begin{align}
		Q^*_{\leq k}(0) &\leq Q^*_{\leq k'}(0) \\
		Q^*_{\leq k}(1) &\geq Q^*_{\leq k'}(1)\;,
	\end{align}
	which further implies that, unless $Q^*_{\leq k}$ and $Q^*_{\leq k'}$ are equal, there is at least one point $\beta$ such that for all $\alpha'>\beta$ we have $Q^*_{\leq k'}(\alpha')\leq Q^*_{\leq k}(\alpha')$ as a consequence of $Q^*_{\leq k}$ and $Q^*_{\leq k'}$ being non-increasing, from which follows the proof.
	If indeed $Q^*_{\leq k}$ and $Q^*_{\leq k'}$ were equal, both binary vectors $\bx$ and $\bx'$ with $\norm{\bx}_1=k$ and $\norm{\bx'}_1=k'$ would be optimal, from which the proof still follows.
\end{proof}

\section{Datasets}
\label{sec:datasets}

In the following, we will briefly describe each data set used in \cref{sec:experiments}.

\begin{itemize}
	\item \texttt{mnist}~\citep{lecun_mnist_2010} contains $28\times 28$ gray-scale images of handwritten digits.
	Each feature (\ie, each pixel) can take integer values from 0 (black) to $255$ (white).
	We divided these values by $255$ to rescale the features to the range $[0,1]$.
	There are ten classes, one for each digit.
	\item \texttt{ionosphere}~\citep{sigillito_classification_1989,dua_uci_2017} contains measurements of electrons in the ionosphere captured by $16$ antennae in north-east Canada.
	The resulting $34$ features take values in the range $[-1,1]$.
	The binary label indicates the presence of evidence of certain structures in the ionosphere.
	\item \texttt{waveform} is a synthetic data set first introduced in \cite{breiman_classification_1984}.
	It contains short time series of length 21, each containing a random linear combination of two of three triangular base waves with added Gaussian noise.
	The $\binom{3}{2}=3$ combinations of two out of three base waves provide the class label.
	\item \texttt{madelon} consists of 5-dimensional points sampled around the corners of a hypercube, with each corner randomly representing one of two classes.
	In addition, 15 linear combinations of these five features as well as 480 random irrelevant features (``probes'') without predictive power are included, leading to 500 features in total.
	\item \texttt{synth\_10} is another synthetic data set with $n=10$ features and a binary label that indicates if a linear combination of a subset of four specific features is above a fixed threshold.
	We generated the data set by first choosing $d_{\text{inf}}=4$ indices of informative features $\cI$ uniformly at random from $[n]$ with $n=10$.
	We then sampled two random correlation matrices $\bC_{\text{inf}}$ and $\bC_{\text{rest}}$ with dimensions $d_{\text{inf}}\times d_{\text{inf}}$ and $d_{n-\text{inf}}\times d_{n-\text{inf}}$ respectively, using the algorithm under section 3.2 of \cite{lewandowski_generating_2009} with $\beta=1$.
	Next, we sample i.i.d. $\mu_i\sim\mathcal{N}(0,10)$ and $\sigma_i\sim\exp(\mathcal{N}(0,1))$ for $i\in [n]$, such that $\bmu_{\text{inf}}=(\mu_i)_{i\in\cI}$ and $\bmu_{\text{rest}}=(\mu_j)_{j\in [n]\backslash\cI}$, and $\bsigma_{\text{inf}}$ and $\bsigma_{\text{rest}}$ respectively.
	From this, we obtain covariance matrices $\bSigma_{\text{inf}}=\bsigma_{\text{inf}}\bsigma_{\text{inf}}^{\top}\odot\bC_{\text{inf}}$ and $\bSigma_{\text{rest}}$ analogously.
	Finally, we sample a data point $\bx$ with $\bx_{\cI}\sim\mathcal{N}(\bmu_{\text{inf}}, \bSigma_{\text{inf}})$ and $\bx_{[n]\backslash\cI}\sim\mathcal{N}(\bmu_{\text{rest}}, \bSigma_{\text{rest}})$.
	We generate the labels by sampling $\bw\in\mathbb{R}^{d_{\text{inf}}}$ with $w_i\sim\mathcal{N}(0,1)$ i.i.d. for all $i\in [d_{\text{inf}}]$.
	Then, for each data point $\bx$ we set the label to $y=0$ if $z:=\bw^{\top}\bx$ is below its mean value $\mathbb{E}_{\bx}[z]$, and to $y=1$ otherwise, which yields, in expectation, an equal class distribution.
	\item \texttt{synth\_100} is generated using the same procedure as \texttt{synth\_10}, but with $n=100$ and $d_{\text{inf}}=10$.
\end{itemize}

\end{document}